\documentclass[prl,twocolumn,showpacs,preprintnumbers,amsmath,amssymb,superscriptaddress,nofootinbib,english]{revtex4-1}

\usepackage{graphicx}
\usepackage{dcolumn}
\usepackage{bm}
\usepackage{epsfig}
\usepackage{graphicx}
\usepackage{hyperref}
\usepackage[usenames]{color}
\usepackage{url}
\usepackage{mathrsfs}
\hypersetup{
    colorlinks=true,
    linkcolor=red,
    citecolor=blue,
}

\newcommand{\tc}{\textcolor{black}}

\newcommand{\remove}[1]{}

\def\ie{{\frenchspacing\it i.e.}}
\def\eg{{\frenchspacing\it e.g.}}

\def\be{\begin{equation}}
\def\ee{\end{equation}}
\def\ba{\begin{eqnarray}}
\def\ea{\end{eqnarray}}

\def\lnk{\kappa}
\def\lnkp{\kappa'}
\def\d{\rm d}

\def\p{{\rm prior}}
\def\fid{{\rm fid}}

\newcommand{\planck}{\textit{Planck}}

\begin{document}

\title{Reconstructing primordial power spectrum using \planck{} and SDSS-III measurements}

\author{Xin Wang}
    \email{xinwang@physics.ucsb.edu}
    \affiliation{National Astronomy Observatories, Chinese Academy of Science, Beijing, 100012, 
    P.R.China}
    \affiliation{School of Astronomy and Space Science, Nanjing University, Nanjing 210093, China}

\author{Gong-Bo Zhao}
    \email{Gong-Bo.Zhao@port.ac.uk}
    \affiliation{National Astronomy Observatories, Chinese Academy of Science, Beijing, 100012, 
    P.R.China}
    \affiliation{Institute of Cosmology and Gravitation, University of Portsmouth, Portsmouth, PO1 
    3FX, UK}

\begin{abstract}
  We develop an accurate and efficient Bayesian method to reconstruct the primordial power spectrum 
  in a model-independent way, and apply it to the latest cosmic microwave background measurement 
  from \planck{} mission, and the large scale structure observation of SDSS-III BOSS (CMASS) sample, 
  combined with the type Ia supernovae sample (SNLS 3-year) and the measurements of baryon acoustic 
  oscillations from SDSS-II, 6dF, and WiggleZ survey. We confirm that the scale-invariant primordial 
  power spectrum is strongly disfavored, and a model with suppressed power on horizon scales is 
  supported by current data.  We also find that a modulation on scales   
  $5\times10^{-4}~\textrm{Mpc}^{-1} \lesssim k \lesssim 0.01~\textrm{Mpc}^{-1}$ is mildly preferred 
  at $2\sigma$ confidence level, whose origin needs further investigation.
\end{abstract}


\maketitle

  The reconstruction of the primordial power spectral amplitude $A_s(k)$ directly from cosmological 
  observations provides the key to understanding the physics of the early universe. A 
  scale-dependent $A_s(k)$, if confirmed, clearly supports the inflation paradigm, and thus various 
  inflationary models can be differentiated by the specific scale-dependence, \eg, the large-scale 
  modulations and the small-scale features. This theoretical significance has motivated many efforts 
  in the literature to reconstruct $A_s(k)$ either parametrically (more often using a power-law 
  parametrization), or non-parametrically. 

  Non-parametric reconstruction of $A_s(k)$ is receiving more and more attention since the result 
  can be largely immune to theoretical bias because no {\it ad hoc} functional form of $A_s(k)$ 
  needs to be assumed, which is inevitable in parametric approaches. However, accurate and efficient 
  non-parametric methods are in general difficult to design and implement because it needs to 
  satisfy some requirements. For instance, (I) it should allow a sufficient number of degrees of 
  freedom (d.o.f.'s) to find significant large- and small-scale features in $A_s(k)$, if there are 
  any, (II) it must not over-fit data, \ie, avoidance of fitting noise, (III) it should include
  reconstruction error analysis, favorably in Bayesian nature, (IV) other cosmological parameters 
  can be varied simultaneously to account for parameter degeneracies and (V) it should be applicable 
  to any kinds of data, including geometrical indicators, \eg{}, baryon acoustic oscillations (BAO), 
  type Ia supernovae (SNIa).

  A lot of methods have been proposed in the spirit of binning \cite{binning}, \ie, fitting constant 
  values of $A_s(k)$ in several $k$ bins to data. These methods can in principle satisfy (III) and 
  (IV), but it is difficult to have sufficient number of bins due to parameter degeneracies. Direct 
  inversion methods \cite{cosmic_inversion} can solve this problem, but they in general do not 
  satisfy (III-V). Fitting principle components is less affected by parameter degeneracies 
  \cite{Leach:2005av}, but zeroing the poorly-constrained high frequency modes, which is practically 
  necessary when employing the Markov Chain Monte Carlo (MCMC) method, can bias the reconstruction 
  result in a non-trivial way \cite{Huterer:2002hy}. The multi-resolution methods, including the 
  direct wavelet expansion \cite{wavelet}, are promising in feature detection, yet it is difficult 
  to avoid under- or over-fitting data. The methods with a penalty term in likelihood calculation 
  are designed to avoid data over-fitting, but using a few fitting nodes with interpolation 
  \cite{Sealfon:2005em} might artificially smooth out signals, hence violates (I), while the method 
  developed in \cite{Gauthier:2012aq} and applied in \cite{planck:inflation} can
  hardly satisfy (IV) and (V).

  In this work, we employ the correlated prior method recently developed for dark energy equation of 
  state reconstruction \cite{cpz1,cpz2,cpz3} to reconstruct the primordial power spectrum using 
  mainly \planck{} 2013 \cite{planck:cosmology} and SDSS-III BOSS measurements \cite{boss}. This
  non-parametric Bayesian reconstruction technique satisfies rigorously all the requirements (I-V). 

  Suppose $A_s(k)$ is a Gaussian random field with a covariance described by a correlation function,
    \be \label{eq:xi}
        \xi (|\lnk - \lnkp|) \equiv \left\langle [A_s(\lnk) - A_s^{\rm fid}(\lnk)][A_s(\lnkp) - 
        A_s^{\rm fid}(\lnkp)] \right\rangle
    \ee
  where $\lnk\equiv \ {\rm ln} \ k$. Discretizing $A_s(\lnk)$ into $N_{\lnk}$ bins in the range of 
  $[\lnk_{\rm min}, \lnk_{\rm max}]$, one can calculate the $\{i,j\}$ component of the covariance 
  matrix for the correlated prior,
    \be \label{eq:Cij}
        C_{ij} = \frac{1}{\Delta^2} \int_{\lnk_i}^{\lnk_i+\Delta}{\d}\lnk 
        \int_{\lnk_j}^{\lnk_j+\Delta}{\d}\lnk' ~ \xi (|\lnk - \lnkp|)
    \ee
  where $\Delta$ is the bin width. We adopt the CPZ form for the correlation function due to its 
  relatively simple behavior and transparent dependence on its parameters \cite{cpz2}, \ie, 
  $\xi(\delta\lnk)=\xi(0)/[1+(\delta\lnk/\lnk_c)^2]$, where $\lnk_c$ determines the correlation 
  length and the amplitude $\xi(0)$ sets the strength of the prior. The variance of the mean $A_s$ 
  over all the bins simply follows from Eq.~(\ref{eq:Cij}) when taking $i=j$ and $\Delta$ to be the 
  entire $\lnk$ interval, and in the limit of $\lnk_c\ll\lnk_{\rm max}-\lnk_{\rm min}$, this 
  variance can be calculated as,
    \be
        \sigma^2_{\bar{P}}=\int_{\lnk_{\rm min}}^{\lnk_{\rm max}} \int_{\lnk_{\rm min}}^{\lnk_{\rm 
        max}}\frac{{\d}\lnk{\d}\lnk'~\xi (\lnk - \lnkp)}{(\lnk_{\rm max}-\lnk_{\rm 
        min})^2}\simeq\frac{\pi\xi(0){\lnk_c}}{\lnk_{\rm max}-\lnk_{\rm min}}
    \ee
  A strong prior (large $\lnk_c$, small $\sigma^2_{\bar{P}}$, hence small $\xi(0)$) results in small 
  variance of the reconstruction, but may bias the reconstructed model when the true model is in 
  tension with the peak of the prior. Using a much weaker prior can avoid biasing the result, but it 
  inevitably leads to a very noisy reconstruction, in other words, over-fit data. To find reasonable 
  values for the prior, we perform tests on an ensemble of inflationary models with different 
  potentials, and we find that taking $\sigma^2_{\bar{P}}=0.072, \lnk_c=0.6$ yields accurate 
  reconstruction with negligible bias.  We take this prior to be the `standard' prior. To be 
  conservative, we also consider a `weak' prior with $\sigma^2_{\bar{P}}=0.144, \lnk_c=0.6$ in case 
  that the true $A_s$ is not covered by the suite of models we use for the bias test.   

  The correlated prior for the model is then
    \be \label{eq:prior}
        \mathcal{P}_{\p}\propto {\rm exp}\left[-\left(\mathbf{A_s}-\mathbf{A_s^{\fid}}\right)^T 
        \mathbf{C}^{-1} \left(\mathbf{A_s}-\mathbf{A_s^{\fid}}\right)/2\right]
    \ee
  To incorporate this prior with MCMC, we minimize the total posterior $\chi^2\equiv\chi_{\rm 
  data}^2+\chi_{\p}^2$ where $\chi_{\p}^2=-2 \ {\rm ln}\ \mathcal{P}_{\p}$. As discussed in 
  \cite{cpz2}, the correlated prior can effectively gauge the flat directions in parameter space, 
  which enables MCMC calculations to converge even for a large number of bins. This allows for a 
  high-resolution reconstruction of $A_s(\lnk)$ without over-fitting data since the correlated prior 
  penalizes the high-frequency modes in such a fashion that the oscillatory modes with low 
  significance are effectively washed out, while the features with high significance, including 
  those sharp ones, are not affected by the prior.   

  To avoid biasing the result by assuming any fiducial model $A_s^{\rm fid}$ in 
  Eq.~(\ref{eq:prior}), we marginalize over it following \cite{cpz2,cpz3} to take local average of 
  the neighboring trial bins within a range of $\Delta\lnk=\lnk_c=0.6$. We have checked our result 
  by adopting another marginalization method of panelizing ${\rm d}A_s/{\rm d\lnk}$ instead 
  \cite{cpz3}, and found a consistent result.  

  \begin{figure}[tbp]
    \includegraphics[scale=0.15]{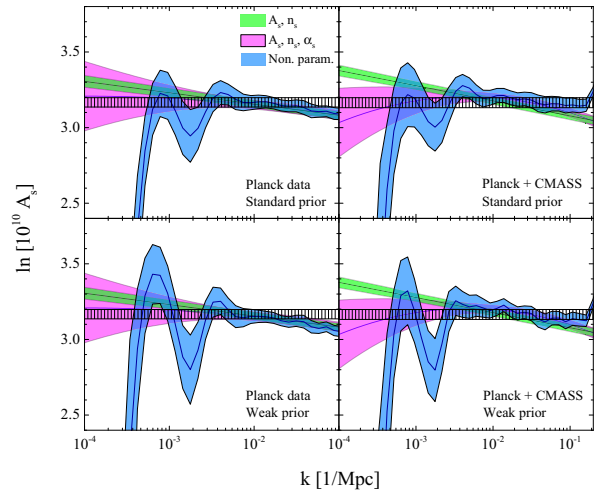}
    \caption{The best fit (solid curves) and 68\% CL error (shaded bands) of the reconstructed 
    primordial power spectrum using the power law parametrization (green and purple shaded) and 
    free-form with correlated priors (blue shaded). Different datasets and priors are employed as 
    illustrated in the legends. The horizontal bands with patterns show the 68\% CL constraint on 
    the HZ model.}\label{fig:As}
  \end{figure}

  In practice, we fit ln$A_s(\kappa)$ to data since it is closer to Gaussian distribution. We 
  approximate ln$A_s(\kappa)$ using 40 bins $\mathcal{A_S} \supset \{ {\rm ln} A_s({\lnk_i}),\ 
  i=1,...,40\}$, spaced uniformly in $\lnk$ in the range of $[{\rm ln}10^{-4},{\rm ln}0.3]$, to 
  cover the range of observables we use. The bin width is sufficiently small compared to the 
  correlation length, thus the prior largely wipes out the dependence on the choice of binning. For 
  comparison, we also fit the usual power law model to data, namely,
    \be \label{eq:powerlaw}
        {\rm ln}A_s(k)={\rm ln} A_s+(n_s-1)~{\rm ln}(k/k_0)+\frac{\alpha_s}{2}~{\rm ln}(k/k_0)^2
    \ee
  where $A_s, n_s$ and $\alpha_s$ are constants and $k_0$ is the pivot scale of $0.05$ Mpc$^{-1}$.  
  So in this case $\mathcal{A_S}\supset\{ {\rm ln}A_s, n_s, \alpha_s\}$.

  \begin{figure*}[tbp]
    \includegraphics[scale=0.2]{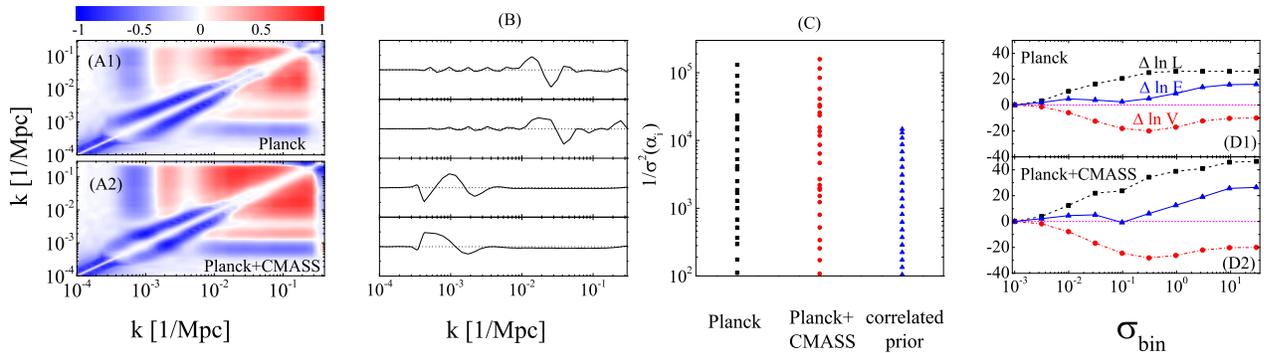}
    \caption{(A1): The correlation matrix among the $\lnk$ bins obtained in MCMC using \planck{} 
    data, subtracted off that of the weak correlated prior; (A2): same as (A1) but for 
    \planck{}+CMASS data; (B): selected eigen-modes of the covariance matrix obtained using 
    \planck{}+CMASS data with a weak correlated prior; (C): The eigen-values of the covariance 
    matrix obtained using \planck{}, \planck{}+CMASS data with a weak correlated prior, and that of 
    the weak correlated prior itself; (D1,2): The improved ln$L$ of the binned model with a weak 
    correlated prior (black dashed); The logarithmic fractional volume ln$V$ of the sampled 
    parameter space (red dash-dot); The logarithmic evidence (blue solid). All three curves are for 
    the residue with respect to that of the constant $A_s$ model as a function of the diagonal prior 
    $\sigma_{\rm bin}$. See text for more details. (D1) and (D2) are for \planck{} and 
    \planck{}+CMASS data respectively.}\label{fig:PCA}
  \end{figure*}

  We apply our method to a joint dataset of the latest cosmological observations. The cosmic 
  microwave background (CMB) and galaxy power spectrum have direct information for $A_s(\lnk)$ and 
  therefore we use the first year CMB measurement from \planck{} satellite\cite{planck:cosmology} 
  and the 3D galaxy power spectrum of SDSS-III BOSS DR9, the CMASS sample \cite{boss}. We model the 
  galaxy bias and the redshift space distortion using the approach developed in \cite{Cole:2005sx} 
  and applied to CMASS in \cite{Zhao:2012xw}. We also include other measurements to constrain the 
  background cosmology to break parameter degeneracies. We use the BAO measurements from SDSS-II 
  \cite{sdss2}, 6dF \cite{6df} and WiggleZ survey \cite{wigglez}, and the SNIa sample of SNLS 3-year 
  \cite{Conley:2011ku}. Note that we didn't combine the $H_0$ measurement in \cite{Riess:2011yx} 
  because of its tension with \planck{} data. Given this joint dataset, we use MCMC  
  \cite{Lewis:2002ah} to sample the parameter space ${\bf P} \equiv (\omega_{b}, \omega_{c}, 
  \Theta_{s}, \tau, \mathcal{A_S},\mathcal{N})$ where $\omega_{b}$ and $\omega_{c}$ are the baryon 
  and cold dark matter densities, $\Theta_{s}$ is the ratio of the sound horizon to the angular 
  diameter distance at decoupling, and $\tau$ is the optical depth. We also include and marginalize 
  over $\mathcal{N}$, which represents the 14 nuisance parameters involved with the \planck{} CMB 
  likelihood and another 2 accounting for the calibration uncertainty in measuring the intrinsic SN 
  luminosity. A modified version of {\tt CAMB} \cite{CAMB} is used to calculate the observables.  
  Note {\emph all} the above parameters are simultaneously varied in our reconstruction.

  The reconstruction result is shown in Fig.~\ref{fig:As}. The blue shaded regions on top layers in 
  four panels illustrate the 68\% confidence level (CL) uncertainties of our reconstruction, while 
  the solid curves inside the bands show the best fit $\mathcal{A_S}$ models. The reconstructions 
  using different correlated priors (standard and weak) and diverse data combinations (\planck{} and 
  \planck{}+CMASS) are displayed separately. Here the datasets of SNIa and BAO are always utilized 
  throughout our analysis. In all cases, we can identify a significant signal of the lack of power 
  on large scales ($k\lesssim 5\times10^{-4}$ Mpc$^{-1}$), which is also apparent in the \planck{} 
  CMB data.  Interestingly, we find a sign of modulation on scales $5\times10^{-4}~\textrm{Mpc}^{-1} 
  \lesssim k \lesssim 0.01~\textrm{Mpc}^{-1}$. Adding the CMASS sample makes the modulation slightly 
  less significant, but still obvious. For a comparison, we also show the reconstructions assuming 
  the usual power-law parametrization, \ie, Eq.~(\ref{eq:powerlaw}).  The purple or green shaded 
  bands represent the cases in which the running $\alpha_s$ is fixed to 0 or allowed to vary 
  respectively, with the corresponding best fit models plotted within their bands.  We also did 
  another fit for the Harrison-Zel'dovich (HZ) model, \ie, $n_s=1,~\alpha_s=0$ (shown in horizontal 
  bands with patterns). In all cases, we can see that the power-law reconstructions are in agreement 
  with the free-form ones on scales of $5\times10^{-3}~\textrm{Mpc}^{-1} \lesssim k \lesssim 
  0.1~\textrm{Mpc}^{-1}$. On larger scales $k\lesssim5\times10^{-3}$ Mpc$^{-1}$, the power law 
  reconstructions fail to capture the suppression of power, and the modulation. This is expected due 
  to the lack of d.o.f.'s in the power-law form. Interestingly, from Table I we see that the 
  inclusion of CMASS data changes mean values of the tilt $n_s$ from 0.9653 to 0.9500 ($\alpha_s$ 
  fixed to 0) and from 0.9620 to 0.9477 ($\alpha_s$ float) respectively. Moreover, \planck{}+CMASS 
  data mildly favors a non-zero running at about $2.3\sigma$. The origin of this inconsistency 
  between two datasets is unclear, and asks for further investigation.   

\begin{table}[htdp]
\begin{center}
\begin{tabular}{c|c|c}
\hline\hline   

     &  \multicolumn{2}{c}{Data Combinations} \\  \cline{2-3}
     & \planck{} & \planck{}+CMASS  \\
     \hline
ln$(10^{10} \ A_s)$ & $3.137\pm0.031$ &  $3.181\pm0.033$ \\
     \hline
ln$(10^{10} \ A_s)$ & $3.091\pm0.026$ &  $3.098\pm0.024$ \\
$n_s$ & $0.9653\pm0.0057$ & $0.9500\pm0.0053$ \\
\hline
ln$(10^{10} \ A_s)$ & $3.110\pm0.032$ & $3.139\pm0.031$   \\
$n_s$ & $0.9620\pm0.0063$& $0.9477\pm0.0059$\\
$\alpha_s$ & $-0.0128\pm0.0090$& $-0.0205\pm0.0088$\\
\hline\hline   
\end{tabular}
\caption{The mean and 68\% CL error of the power-law parameters when they vary.}
\end{center}
\label{tab:constraint}
\end{table}%

  To quantify the goodness of fit using different parametrisations, we list the $\chi^2$ for the 
  corresponding best fit models relative to that for the HZ model in Table II. We can see that the 
  HZ model is strongly disfavored in all cases, with the significance ranging from $5.6\sigma$ 
  (\planck{}, $n_s$ float, $\alpha_s$ fixed) to $9.6\sigma$ (\planck{}+CMASS, weak prior). For the 
  power law scenario with \planck{}+CMASS, $\chi^2$ can be reduced by $12.5$ if $\alpha_s$ is 
  allowed to vary, which is consistent with what we see in Table I: a non-zero running is 
  unambiguously preferred by this data combination. If we allow additional d.o.f.'s, \ie, adopting 
  the free-form parametrization, $\chi^2$ can be drastically reduced. For example, with 
  \planck{}+CMASS, $\chi^2$ for the weak prior case is lower than the power law case with free 
  running by $16.1$, which is a $4\sigma$ significance.

\begin{table}[htdp]
\begin{center}
\begin{tabular}{c|c|c|c|c}
\hline\hline   

     &  \multicolumn{2}{c|}{Power Law} &\multicolumn{2}{c}{Free-Form} \\  \cline{2-5}
     & $n_s$ & $n_s,~\alpha_s$ & Standard Prior & Weak Prior \\
     \hline
\planck{} & $-31.5$ & $-32.4$& $-47.6$& $-51.7$ \\
\hline
\planck{}+CMASS & $-64.1$& $-76.6$& $-83.7$& $-92.7$\\
\hline\hline   
\end{tabular}
\caption{The improved $\chi^2$ of the power law and free-form models with respect to the 
Harrison-Zel'dovich model ($n_s=1$) using two datasets (\planck{} and \planck{}+CMASS). }
\end{center}
\label{tab:chi2}
\end{table}%

  To understand this result and confirm that we are not over-fitting data, we perform a principle 
  component analysis and calculate the Bayes factor explicitly. We diagonalise the covariance matrix 
  for the $\mathcal{A_S}$ bins, which is obtained from the posterior distribution, in order to find
  the uncorrelated linear combinations of the bins with all other cosmological parameters 
  marginalized over. Panels (A1, A2) in Fig.~\ref{fig:PCA} show the correlation matrix (the 
  normalized covariance matrix so that all the diagonal terms are $1$) among the bins subtracted off 
  the weak prior correlation matrix for \planck{} and \planck{}+CMASS data respectively.  Note that 
  if data are absent, the correlated prior provides a positive correlation among bins within the
  correlation length, so the prior correlation matrix is block diagonal with positive entries in the 
  off-diagonal terms. When data are added in, this correlation pattern can be changed significantly 
  where data are strong, or not affected much where data are weak. From (A1, A2) we can recognize 
  that data require negative correlation among bins on scales $k\lesssim0.01$ Mpc$^{-1}$, meaning 
  that a large variation of amplitudes on these scales is favored. This is consistent with what we 
  see in Fig.~\ref{fig:As}.  On smaller scales, the correlation is positive, suggesting that 
  amplitudes on these scales behave more coherently, and this is the reason that there is no 
  apparent features seen on these scales in our reconstruction.  The inclusion of CMASS data 
  slightly changes the correlation pattern, \eg, the correlation on quasi-nonlinear scales 
  ($0.1~\textrm{Mpc}^{-1} \lesssim k \lesssim 0.2~\textrm{Mpc}^{-1}$) is more negative, implying a 
  feature on such scales, which is seen in Fig.~\ref{fig:As}.  However, this feature is likely due 
  to systematics, \eg, issues of nonlinearity rather than being physical. 

  Panels (B, C) show the eigen-vectors and eigen-values of the covariance matrix. From panel (B), we 
  see that the well constrained modes modulate on scales where we see features in the 
  reconstruction, and panel (C) quantitatively shows that \planck{} and \planck{}+CMASS can 
  constrain 8 and 10 such modes respectively. 

  Our free-form reconstructions apparently fit data better than the HZ or the power-law models, but 
  the key issue is whether this is ascribed to an over-fit of the data, in other words, whether this 
  fitting improvement can compensate for the increased volume of parameter space. This can be 
  quantified by computing the Bayes factor $E$ within a family of models interpolating smoothly 
  between the weak prior free-form model and the HZ model. We follow \cite{cpz3} and implement this 
  via adding a larger and larger diagonal term to the inverse prior matrix which effectively reduces 
  the variance in each bin. This essentially shifts all the eigen-values by a constant. The Bayes 
  factor $E$ can be estimated as,
    \be
        E \propto VL; ~V=\sqrt{\frac{\det{{\cal C}_{\rm post}}}{\det{{\cal C}_{\rm prior}}}}; ~L 
        =e^{-\chi_{\rm b.f.}^2/2},
    \ee
  where ${\bf {\cal{ C}}}_{\rm prior}$ and ${\bf {\cal{ C}}}_{\rm post}$ denote the prior and 
  posterior covariance matrices respectively, and $\chi_{\rm b.f.}^2$ is the $\chi^2$ for the best 
  fit model given the combination of data and prior. Note that $V$ quantifies the fraction of the 
  parameter space corresponding to the initial prior consistent with data, while $L$ indicates how 
  well the model is capable of fitting the data. Panels (D1, D2) in Fig.~\ref{fig:PCA} show ${\rm 
  ln}L,~{\rm ln}V$ and ${\rm ln}E$ with respect to those in the HZ model as a function of 
  $\sigma_{\rm bin}$ calculated using \planck{} and \planck{}+CMASS data respectively. As we see, 
  $\Delta {\rm ln} E$ is non-negative for all $\sigma_{\rm bin}$, inferring the necessity of 
  free-form reconstructions of $A_s(k)$. Especially when $\sigma_{\rm bin}$ approaches 20, which 
  corresponds to the weak prior model, $\Delta {\rm ln}E$ is $16.1$ and $26.4$ for \planck{} and 
  \planck{}+CMASS respectively. 

  In this letter, we develop a new and robust Bayesian method to reconstruct the primordial power 
  spectrum in a non-parametric way using latest cosmological observations including \planck{} and 
  SDSS-III measurements. We find that the scale-invariant spectrum is strongly disfavored, while a 
  model with suppressed power on large scales ($k\lesssim5\times10^{-4}$ Mpc$^{-1}$) is supported by 
  data. A sign of modulation on scales $5\times10^{-4}~\textrm{Mpc}^{-1}\lesssim k \lesssim
  0.01~\textrm{Mpc}^{-1}$ is also evidenced by current CMB and large scale structure data.  Whether 
  it stems from new physics in the early universe \cite{Feng:2003zua} or some unaccounted 
  systematics can be shed light upon with the upcoming polarization data from \planck{} and future 
  large redshift surveys.
  
  \acknowledgments We thank Wayne Hu, Kazuya Koyama and Levon Pogosian for useful comments and 
  discussions. XW is supported by NAOC. GBZ is supported by University of Portsmouth, and the {\it 
  1000 young talents} program in China.

\end{document}